\documentclass[doublecol]{epl2}
\usepackage{amsmath}
\begin{document}

\title{Coherent spin rotation in the presence of a phonon-bottleneck effect}

\author{L. Chen\inst{} \and I. Chiorescu\inst{}}
\institute{Physics Department and the National High Magnetic Field Laboratory, Florida State University, 1800 E. Paul Dirac Drive, Tallahassee, Florida 32310\\
E-mail: ichiorescu@fsu.edu} 

\date{to appear}
\abstract{
A characteristic of spin reversal in the presence of phonon-bottleneck is the deviation of the magnetization cycle from a reversible function into an opened hysterezis cycle. In recent experiments on molecular magnets (e.g. V$_{15}$ and Ru$_2$), the zero-field level repulsion was sufficiently large to ensure an otherwise adiabatic passage through zero-field and the magnetization curves can be described by using only a phonon-bottleneck model. Here, we generalize the phonon-bottleneck model into a model able to blend the non-adiabatic dynamics of spins with the presence of a non-equilibrium phonon bath. In this simple phenomenological model, Bloch equations are written in the eigenbasis of the effective spin Hamiltonian, considered to be a two-level system at low temperatures. The relaxation term is given by the phonon-bottleneck mechanism. To the expense of calculus time, the method can be generalized to multi-level systems, where the notion of Bloch sphere does not apply but the density matrix formalism is still applicable.\\
EPL, 87 (2009) 57010; 23 September 2009; doi: 10.1209/0295-5075/87/57010
}

\maketitle 

\section{Introduction}
Molecular magnets \cite{1} or less complex systems containing diluted spins \cite{2} have been explored intensively in the recent years for their potential application in information technology. In addition, despite their macroscopic dimensions, crystals containing magnetic molecules show measurable manifestations of fundamental quantum mechanical phenomena at large scale. This is due only to the fact that the magnetic molecules are relatively well separated from each other and their quantum properties are amplified by the large number contained in a tiny monocrystal. Consequently, they show remarkable phenomena, like Quantum Tunneling of the Magnetization (QTM) measured first \cite{3}  in spin $S=10$ systems (Mn$_{12}$, Fe$_8$) and Berry Phase quantum interference \cite{4} (Fe$_8$). This large spin molecular magnets show significant anisotropy barriers protruded by tunneling channels which are generated by a transverse magnetic field or anisotropy terms. A sweeping longitudinal field (that is, parallel with the main spin quantification axis) will force the spin to sequentially visit the tunneling channels and obey the Landau-Zener tunneling mechanism each time. Consequently, distinct jumps are generated in the hysterezis cycle \cite{3} allowing a direct measure of the tunneling gap \cite{5}.

Another type of molecular magnets consist of species having a small total spin, e.g. $S=1/2$ (V$_{15}$ \cite{6}, V$_6$ \cite{7}) or $S=1$ (Fe$_{10}$) \cite{8}. For a two-level system ($S=1/2$) one can have $S_+$, $S_-$ transverse terms generated by the complex structure of the molecules (like Dzyaloshinski-Moriya interactions in V$_{15}$ \cite{9}) leading to a level repulsion in zero field $\Delta$ (similar to a tunneling channel in large spin molecules). Due to the low value of the spin, there is no or low anisotropy barrier between states with opposite projections of $S_Z$. For $\Delta$ large enough to allow the spin to flip adiabatically under swept field, the magnetization cycle is reversible at thermal equilibrium. However, experiments performed in V$_{15}$ and other molecules \cite{9,10} have shown an opening of the hysterezis cycle despite the absence of a barrier. Such phenomena have been interpreted within the frame of the phonon-bottleneck model which predicts a hysterezis of spin temperature in the presence of a fast sweeping field. For very small values of the level repulsion $\Delta$, the phonon-bottleneck and the Landau-Zener tunneling phenomena have to be treated together, which is the main aspect of this article.

Molecular systems are tunable, identical, two- or multilevel systems that can be produced in large numbers and therefore are potential candidates for qubit implementation in quantum computing algorithms \cite{10a}. In practice however, spin qubits are very sensitive to decoherence mechanisms, mainly due to hyperfine and dipolar interactions with the background spins, and also to environmental thermal effects (phonon bath). Advanced chemistry techniques allow synthesizing single crystals with a reduced amount of background magnetic elements leading to improved coherence times \cite{2,11}, whereas low temperatures reduce the phonon bath effect in usual experiments. In this paper we analyze the possibility of quenching thermal decoherence effects by operating the spins in phonon bottleneck (PB) conditions.
\section{Modeling}
The phonon-bottleneck phenomenon is generated by an imperfect thermalization of a sample with the surrounding experimental setup. At low temperatures, the number of available phonons is strongly suppressed and the inherent limited sample thermalization with the cryogenic bath will delay the return to equilibrium of sample's phonon bath. In the case of a magnetic sample, phonons are absorbed and re-emitted resonantly by the spin system. The timing of such a process, in the presence of a swept magnetic field, generates a bottleneck (strong absorption) followed by a phonon avalanche (delayed strong re-emission) \cite{12}.

At low temperatures, phonons have a very low heat capacity \cite{12} ensuring that the lattice and the spins are at the same temperature $T_S$, which can be different however from the bath (cryostat) temperature $T$. As shown below, cycling of an external magnetic field can induce phonon bottleneck and a delay (hysteresis) in the dynamics of the spin temperature, which appears experimentally as a hysterezis in the magnetic moment of a single crystal \cite{9,10}. In Ref.~\cite{12} a PB model based on the detailed phonon balance is given and its implementation for the case of molecular magnets is presented in \cite{9,10}. In the case of a two-level spin system (e.g. $S=1/2$), the model requires two fit parameters, $\alpha$ and $\Delta$  . The first one is a phenomenological parameter related, among other, to sample's thermalization and the second one is the level repulsion (or the tunneling gap) in zero field. Other theoretical models have been discussed as well in the literature \cite{13}; they are able to simulate magnetic hysterezis but are less adapted to study the relation between sample thermalization and a phonon bath out of equilibrium. Also, recent microscopic modeling solving numerically the density matrix equation does give good quantitative description of experimental results \cite{14}.

A more intuitive model is discussed below and it consists in a modification of the well-known Bloch equations used in spin dynamics, to account phenomenologically for both the Landau-Zener process (not accounted for in the original model \cite{12}) and the PB effect.

Usually, a two-level system is described by the effective Hamiltonian $H_{eff}=-\varepsilon\sigma_z/2-\Delta\sigma_x/2$ where $\sigma_{z,x}$ are the Pauli spin matrices, $\Delta$ is the level repulsion and  $\varepsilon=g\mu_BH$ is the spin bias ($\mu_B$ is the Bohr magneton and $H$ is the external magnetic field, always $\|$ to the $z$ axis of $H_{eff}$). The resulting energy separation can be written as $\Delta_H= \sqrt{(\Delta^2+\varepsilon^2)}$ which is also the energy $h\nu =\Delta_H$ of phonons responsible for spin thermalization. In the presence of a phonon bath in equilibrium, the spin dynamics can be described by the well-known phenomenological Bloch equations, particularly when a spin resonance is driven by a microwave excitation. The Bloch equations are equally useful when studying the effect of a ramping field on the dynamics of an isolated spin. For instance, when sweeping non-adiabatically $\varepsilon$  from $-\infty$ to $+\infty$, the final value of $\langle S_z\rangle$ obeys the Landau-Zener \cite{15} theory:
$P_{LZ}=1-\exp(-\frac{\pi\Delta^2}{2\hbar d\varepsilon/dt})$,
where $P_{LZ}$ is the probability for $\langle S_z\rangle$ to remain in the original state (\emph{up} or \emph{down}).

In analogy to the classical case of magnetic resonance ($\Delta=0$), we will discuss first the Bloch equations in the diagonal basis of $H_{eff}$ and how they can include the effect of phonon bottleneck.

The magnetization components are given by $M_i=\langle\sigma_i\rangle=Tr(\rho\sigma_i)$, $i=x,y,z$ where $\sigma_i$ are the Pauli matrices and $\rho$is the density matrix. For a two level system one gets:
\begin{equation}\label{eq1}
\left\{\begin{array}{ll}
M_x=Tr\begin{pmatrix}\rho_{11}&\rho_{12}\\ \rho_{21}&\rho_{22}\end{pmatrix}\begin{pmatrix}0&1\\1&0\end{pmatrix} =\rho_{12}+\rho_{21}
\\M_y=Tr\begin{pmatrix}\rho_{11}&\rho_{12}\\ \rho_{21}&\rho_{22}\end{pmatrix}\begin{pmatrix}0&-i\\i&0\end{pmatrix} =i(\rho_{12}-\rho_{21})\\
M_x=Tr\begin{pmatrix}\rho_{11}&\rho_{12}\\ \rho_{21}&\rho_{22}\end{pmatrix}\begin{pmatrix}1&0\\0&-1\end{pmatrix} =\rho_{11}-\rho_{22}
\end{array}\right..
\end{equation}
The magnetization dynamics is governed by Liouville equation $\partial\rho/\partial t=-i/\hbar [H(t),\rho]$  which can be written as
\begin{multline*}
i\hbar\begin{bmatrix}\dot{\rho}_{11}&\dot{\rho}_{12}\\\dot{\rho}_{21}&\dot{\rho}_{22}\end{bmatrix}
=\\\begin{bmatrix}-\frac{\Delta}{2}(\rho_{21}-\rho_{12})&-\varepsilon\rho_{12}-\frac{\Delta}{2}(\rho_{22}-\rho_{11})\\
-\varepsilon\rho_{21}-\frac{\Delta}{2}(\rho_{11}-\rho_{22})&-\frac{\Delta}{2}(\rho_{21}-\rho_{12})\end{bmatrix}
\end{multline*}
By evaluating  $\dot{\rho}_{11}\pm\dot{\rho}_{22}$ and $\dot{\rho}_{12}\pm\dot{\rho}_{21}$ one obtains the Bloch equations (with the addition of terms depending on $\Delta$ in absence of any relaxation or dephasing terms):
\begin{equation}\label{eq2}
\left\{\begin{array}{ll}
\dot{M}_x=\frac{\varepsilon}{\hbar}M_y\\
\dot{M}_y=-\frac{\varepsilon}{\hbar}M_x+\frac{\Delta}{\hbar}M_z\\
\dot{M}_z=\frac{\Delta}{\hbar}M_y
\end{array}\right..
\end{equation}
Energy and phase relaxation are described phenomenologically by additional decay terms, characterized by times $T_1$ and $T_2$ respectively. To this purpose, one has first to rotate Equations \eqref{eq2} into the eigenbasis of the spin Hamiltonian to properly add the energy relaxation term. The level repulsion term $\Delta\sigma_x/2$ is equivalent with an applied field along $x$-axis which combines with the applied bias $\varepsilon$ along $z$-axis to give the total magnetic field and the quantification axis of the eigenbasis. It is therefore necessary to rotate the $xyz$ coordinates around the $y$-axis by an angle $\theta=\arctan(\Delta/\varepsilon)$. In the new $x'y'z'$ coordinates, the averaged spin components (or magnetization) are:
\begin{equation}\label{eq3}
\left\{\begin{array}{l}
M_{x'}=M_x\cos\theta-M_z\sin\theta\\
M_{y'}=M_y\\
M_{z'}=M_z\cos\theta-M_x\sin\theta
\end{array}\right..
\end{equation}
Time derivatives of the above set and Equation \eqref{eq2} lead to a new set of motion equations
\begin{equation}\label{eq4}
\left\{\begin{array}{l}
\dot{M}_{x'}=\frac{\sqrt{\varepsilon^2+\Delta^2}}{\hbar}M_{y'}-M_{z'}\dot{\theta}\\
\dot{M}_{y'}=-\frac{\sqrt{\varepsilon^2+\Delta^2}}{\hbar}M_{x'}\\
\dot{M}_{z'}=M_{x'}\dot{\theta}
\end{array}\right..
\end{equation}
One can introduce in Equations \eqref{eq4} the phenomenological energy relaxation term along $z'$-axis and a dephasing term in the $x'y'$ plane. The two processes are assumed to induce exponential decays, characterized by the relaxation time $T_1$ and the dephasing time $T_2$, respectively:
\begin{equation}\label{eq5}
\left\{\begin{array}{l}
\dot{M}_{x'}=\frac{\sqrt{\varepsilon^2+\Delta^2}}{\hbar}M_{y'}-M_{z'}\dot{\theta}-\frac{M_{x'}}{T_2}\\
\dot{M}_{y'}=-\frac{\sqrt{\varepsilon^2+\Delta^2}} {\hbar}M_{x'}-\frac{M_{y'}}{T_2}\\
\dot{M}_{z'}=M_{x'}\dot{\theta}+\frac{M_{z'}^{eq}-M_{z'}}{T_1}
\end{array}\right..
\end{equation}
where $M_{z'}^{eq}$ is the magnetization value at thermal equilibrium. This set of equations, and in particular the energy relaxation term, would correctly describe a typical spin-phonon interaction with the phonon bath at equilibrium. To describe correctly the particular case of a phonon-bottleneck driven process, the $T_1$ term needs to be modified, as discussed below.

\begin{figure}
  \includegraphics[bb=6 63 289 148,width=\columnwidth]{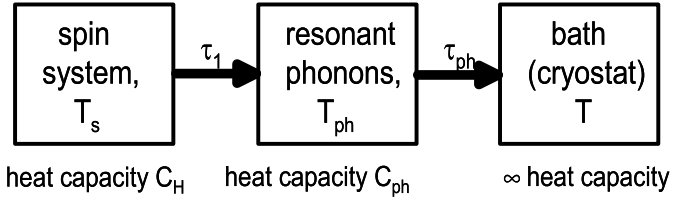}
  \caption{Illustration of the spin-phonon-bath coupling mechanism: spins at temperature $T_s$ and heat capacity $C_H$ are coupled to phonons at temperature $T_{ph}$ and heat capacity $C_{ph}<<C_H$, the energy transfer taking place with a rate $1/\tau_1$. Phonons are coupled to the cryostat at fixed temperature T, their relaxation rate being $1/\tau_{ph}$. At very low temperatures, $T_s=T_{ph}$ and the spin-phonon system relaxes toward T with a rate $C_{ph}/(C_H\tau_{ph})$ (see text).}\label{fig1}
\end{figure}

Even at very low temperatures, the magnetic molecules are still coupled with the thermal heat bath and the spin system will relax towards equilibrium by exchanging energy with the phonons. Only the phonons having the same frequency as the resonant frequency of the spins $h\nu =\Delta_H$ can do the energy transfer. The situation is schematically represented in Fig.~\ref{fig1}. The spin system has a characteristic temperature $T_s$ defined as $n_1/n_2=\exp(\Delta_H/k_BT_S)$ with $n_{1,2}$ the out of the equilibrium occupation numbers of the ground and excited spin level respectively. The phonon bath has a temperature $T_{ph}$ defined by the Bose-Einstein distribution $p_{T_{ph}}{\varepsilon}=1/(\exp(\varepsilon/k_BT_{ph})-1)$. The bath temperature T represents the phonon temperature at equilibrium. The number of resonant phonons in an energy window $\Delta\omega$ is \cite{12}
\begin{equation}
n_T=\int_{\Delta\omega}p_T(\hbar\omega)\sigma(\omega)d\omega
\end{equation}
where $\sigma(\omega)d\omega=3\omega^2d\omega/(2\pi^2\nu^3)$ is the number of phonon modes between $\omega$  and $\omega+d\omega$ per volume unit, $\nu$ is the phonon velocity and $\Delta\omega$ is the transition linewidth. As an example, let us consider the case of $V_{15}$ molecule \cite{6}. Taking the typical values $v$=3000 m/s, $T$=0.1 K and $\Delta\omega\approx$500 MHz, we find $n_T$ of the order of about 10$^{-6}$ to 10$^{-8}$ phonons per molecules which is quite low. These phonons are rapidly absorbed by spins such as $T_s=T_{ph}$ and the coupled spin-phonon system relaxes towards the bath temperature $T$. At low temperatures, absorption of resonant phonons generates a hole in phonon distribution between $\omega$ and $\omega+d\omega$, phenomenon known as \emph{phonon-bottleneck} \cite{16} and seen first experimentally in paramagnetic salts \cite{17}. In such cases both the anharmonic phonon-phonon coupling and the phonon thermalization with the cryostat are insufficient to prevent the hole formation in the phonon distribution.
Spin relaxation implies the transfer of energy from the spin system to the bath via phonons and can be treated simply by means of detailed balance equations. The spin population and the phonon number are given quantitatively by the following two differential equations describing the energy relaxation process:
\begin{equation}
\left\{\begin{array}{l}-\frac{dn_1}{dt}=\frac{dn_2}{dt}=P_{12}n_1-P_{21}n_{2}\\ \frac{dn_{T_{ph}}}{dt}=-\frac{(n_{T_{ph}}-n_T)}{\tau_{ph}}=P_{12}n_1+P_{21}n_2\end{array}\right.
\end{equation}
where $\tau_{ph}$ is the relaxation time of phonons, $P_{12,21}$ are the transition probabilities between the two spin levels and are directly proportional to the available number of phonons($\propto p_{T_{ph}}$ and $p_{T_{ph}}+1$  respectively).

By using the notation
\begin{equation}
1/\tau_1=(P_{12}+P_{21})_{eq}
\end{equation}
for the spin relaxation rate at equilibrium and the variable change (equilibrium quantities are indicated by the subscript eq)
\begin{equation}
x=\frac{n_1-n_2}{n_{1eq}-n_{2eq}},\ y=\frac{p_{T{ph}}-p_T}{p_T-1/2}
\end{equation}
the balance equations can be rewritten as \cite{12}
\begin{equation}\label{eq10}
\left\{\begin{array}{l}dx/dt=(1-x-xy)/\tau_1\\dy/dt=-y/\tau_{ph}+bdx/dt\end{array}\right.
\end{equation}
Here $b=\frac{C_H}{C_{ph}}=\frac{n_1+n_2}{\sigma(\omega)\Delta\omega}\tanh^2(\frac{\hbar\omega}{2k_BT})$
with $C_H$ and $C_{ph}$ spin and phonon heat capacity respectively \cite{12}. Solutions of the system \eqref{eq10} are discussed in \cite{18} for different values of $\tau_{ph}$ and $b$. Generally, in the vicinity of equilibrium ($x\rightarrow1$, $y\rightarrow0$), the spin relaxation is described by an exponential with a characteristic time $\tau_b=\tau_1+(b+1)\tau_{ph}$. Since at very low temperatures $C_H\gg C_{ph}$ and $b\gg1$, $\tau_b$ reduces to $\tau_H=b\tau_{ph}$. For the very same reason, $T_s\approx T_{ph}$ \cite{12}, such that $y=\frac{1}{x}-1$
Therefore, the second equation in \eqref{eq10} becomes
\begin{equation}\label{eq12}
\frac{dx}{dt}=\frac{1}{b\tau_{ph}}\frac{1-x}{x}
\end{equation}
representing the relaxation law of the spin system in the presence of phonon-bottleneck phenomena, in the vicinity of equilibrium. The relaxation time $\tau_H$ depends on the applied magnetic field via the $\Delta_H(\varepsilon)$ dependence and is given by
\begin{equation}\label{eq13}
\tau_H=\frac{\alpha}{\Delta_H^2}\tanh^2(\frac{\Delta_H}{2k_BT})
\end{equation}
with $\alpha=2\pi^2\hbar^2\nu^3N\tau_{ph}/(3\Delta\omega)$ where $N$ is the spin density.

Equation \eqref{eq12} allows modifying accordingly the phenomenological relaxation term in the Bloch equations \eqref{eq5} which was valid only for an equilibrated phonon bath. In the case of phonon bottleneck, eqs. \eqref{eq5} can than be rewritten as
\begin{equation}\label{eq14}
\left\{\begin{array}{l}
\dot{M}_{x'}=\frac{\Delta_H}{\hbar}M_{y'}-M_{z'}\dot{\theta}-\frac{M_{x'}}{T_2}\\
\dot{M}_{y'}=-\frac{\Delta_H}{\hbar}M_{x'}-\frac{M_{y'}}{T_2}\\
\dot{M}_{z'}=M_{x'}\dot{\theta}+\frac{M_{z'}^{eq}}{M_{z'}}\frac{M_{z'}^{eq}-M_{z'}}{\tau_H}\end{array}\right..
\end{equation}
\section{Results and Discussion}
The modified Bloch equations \eqref{eq14} and eq. \eqref{eq13} allow simulating the spin motion during a field sweep in the presence both phonon-bottleneck and Landau-Zener mechanisms (PB-LZ). The obtained hysterezis cycles do compare very well with the experimental ones in the case of large $\Delta$ (that is, without Landau-Zener non-adiabatic transitions), similarly to the PB-only method used in Refs.~\cite{6,9,10}.

\begin{figure}
  \includegraphics[bb=135 345 345 543,width=\columnwidth]{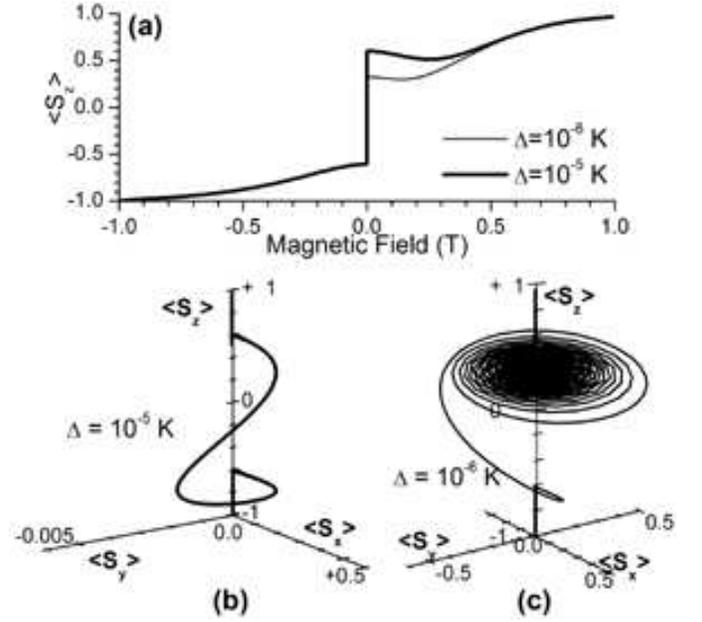}
  \caption{(a) Half hysterezis cycles (field ramped up) for two values of $\Delta$ corresponding to almost adiabatic (thick line) and non-adiabatic (thin line) spin reversal. Due to the PB effect the spin is far from thermal equilibrium. (b) and (c) Spin dynamics in the two situations: adiabatic reversal almost contained in the xz plane and large spin oscillations due to the LZ effect.}\label{fig2}
\end{figure}

To evaluate the effect of the Landau-Zener mechanism, magnetization cycles are calculated for values of $\Delta$ corresponding to non-adiabatic and up to almost adiabatic conditions. Figure~\ref{fig2}a shows $\langle S_z\rangle$ during half-cycles (field up-ramping) with the spin initially aligned to a negative field. The temperature is 0.3~K, magnetic field ramp rate is 0.1 T/s, attainable in current experimental conditions,  $\alpha$=1~sK$^2$ and $\Delta$=10$^{-6}$ K and 10$^{-5}$ K. Such values for $\alpha$ simulate an intermediate sample thermalization (compared with $\alpha\sim$1-3 sK$^2$ in the experiments performed on Ru$_2$ \cite{10} or 0.09/130 sK$^2$ for good/bad thermalization respectively, in the case of the V$_{15}$ experiments \cite{9}). The dephasing time $T_2$ is large (0.1 ms), so the simulations can evidentiate the spin coherent dynamics for a field interval up to 10$^{-5}$ T.

The curves in fig.~\ref{fig2}a give $\langle S_z\rangle$ for negative fields with values in between the zero temperature case and the Boltzmann equilibrium case as a consequence of the phonon-bottleneck effect ($0 < T_s = T_{ph} < T$). The effect is even more visible after passing through zero field, and, for this particular parameters, a valley of negative differential susceptibility is seen (as in the experiments \cite{10}). This is due to an increase in phonon energies as a result of spin de-excitations generating phonons with larger $\hbar\omega$ than those absorbed in lower fields (\emph{phonon avalanche}). The last part of the curves shows a return to equilibrium. The PB-LZ method (14) also gives access to the spin trajectory, as shown in Fig.~\ref{fig2}(b) and (c). In the adiabatic case ($\Delta$=10$^{-5}$ K) the spin trajectory is indeed approximately contained in the $xz$ plane, defined by the field $\Delta$ and $\varepsilon$. For  $\Delta$=10$^{-6}$ K the process is largely non-adiabatic and the LZ mechanism imposes coherent oscillations during the spin flip. They end in a Larmor spin precession around the $z$-axis as imposed by the longitudinal spin bias $\varepsilon$, followed by a phonon-bottleneck and avalanche, as described above (fig.~\ref{fig2}a).

\begin{figure}[t]
  \includegraphics[bb=134 316 347 569,width=\columnwidth]{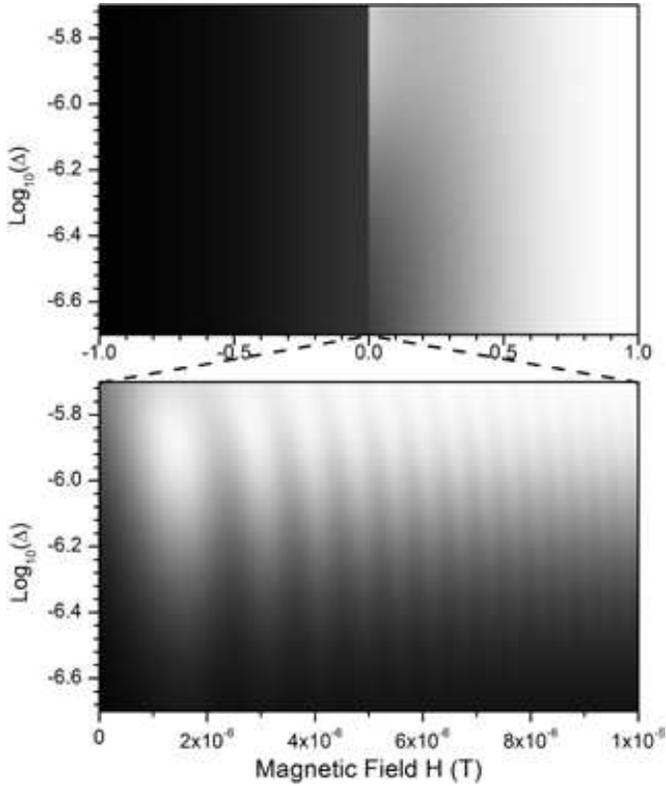}\\
  \caption{Contour plots of half-cycles, with the field ramped from negative to positive values. The $\langle S_z\rangle$ increases gradually from -1 (black) to +1 (white). The zoom shows coherent oscillations induced by the LZ mechanism.}\label{fig3}
\end{figure}

The same mechanisms are demonstrated by the contour plot in Fig.~\ref{fig3}. Here, $\Delta$ is varied continuously between 2$\cdot$10$^{-7}$ K and 2$\cdot$10$^{-6}$ K. Similar half-cycles as in Fig.~\ref{fig2}a are obtained and shown in shades of gray. $\langle S_z\rangle$ varies from -1 (\emph{black}) to +1 (\emph{white}) and one can observe the sudden spin flip in zero field, followed by a phonon-bottleneck and avalanche sequence. The bottom half of Fig.~\ref{fig3} gives a zoom taken at very low and positive magnetic fields. The spin precession is shown as alternating shadings; they gradually vanish at higher fields due to the limited coherence $T_2$ time or at higher values of $\Delta$ where the spin flip enters into the adiabatic regime.
If the field is swept at higher rates and at lower temperatures, the phonon avalanche can be strongly reduced. Thus the lattice phonons and the spins are brought to temperatures much lower than the cryostat one (\emph{adiabatic cooling}). Such behavior has been observed experimentally in the case of the $V_{15}$ molecular magnet \cite{9}. After several field cyclings and for a crystal thermally isolated from the setup, the sample temperature drops significantly and the spin relaxation time is increased from seconds to hours.

In experiments involving ultra-low temperatures and scanning magnetic fields, active cooling of a spin system is therefore possible for a desired amount of time. It is important to note that active cooling experiments have been performed with superconducting quantum bits (\cite{19},\cite{20}). Quenching the lattice at lower temperatures using a PB mechanism could provide an additional knob to improve the active cooling of qubits or of any other quasi-spin system.
\section{Conclusions}
We present a simple phenomenological method to simulate the effect of non-adiabatic Landau-Zener excitations on spin reversal driven by a phonon-bottleneck mechanism. To this purpose, Bloch equations are written in the eigenbasis of the effective spin Hamiltonian, considered to be a two-level system at low temperatures. The relaxation term is given by the phonon-bottleneck mechanism. To the expense of calculus time, the method can be generalized to multi-level systems, where the notion of Bloch sphere does not apply but the density matrix formalism is still applicable.

The authors acknowledge support from the Florida State University, National High Magnetic Field Laboratory (grant IHRP-5059 and DMR-0654118), NSF (grant CAREER DMR-0645408), DARPA (grant HR0011-07-1-0031 ) and the Alfred P. Sloan Foundation.

\end{document}